\documentclass[letterpaper]{article}
\usepackage[T1]{fontenc}
\usepackage{siunitx}
\usepackage{amssymb}
\usepackage{geometry}
\usepackage{setspace}
\usepackage[style = chem-acs]{biblatex}
\usepackage{graphicx}
\usepackage{float}
\newfloat{scheme}{htbp}{los}
\floatname{scheme}{Scheme}
\floatname{chart}{Chart}
\newfloat{graph}{htbp}{loh}

\usepackage{chemformula} 
\usepackage{siunitx}
\usepackage[version = 4]{mhchem} 

\newcommand{\sakuma}[1]{\textcolor{black}{#1}}
\usepackage{authblk}
\author[1]{Ryoko Sakuma*}
\author[1,2]{Koji Sakai}
\author[1]{Hajime Okamoto}
\author[1]{Motoki Asano}
\author[1]{Hiroshi Yamaguchi}

\affil[1]{Basic Research Laboratories, NTT, Inc., Japan}
\affil[2]{NTT Bio-Medical Informatics Research Center, NTT, Inc., Japan}

\title{Optomechanical parametric control of mid-infrared photons\\ via molecular vibrational polariton}
\date{*Email: ryoko.sakuma@ntt.com}

\begin{document}

\maketitle

\begin{abstract}
Controlling mid-infrared (MIR) photons using well-developed telecom photonic platforms would enable new functionalities in molecular and quantum photonics. However, establishing efficient interactions between MIR and telecom photons remains challenging due to their large spectral separation and weak nonlinear coupling. Here, we demonstrate optomechanical control of MIR photons mediated by vibrational polaritons, enabling photon–photon interaction between MIR and telecom fields across distant spectral regions. Using a Fabry--Pérot cavity incorporating a vibrationally active polymer, we observe telecom-driven \sakuma{dissipation enhancement} of MIR photons at $9.5~\mu\mathrm{m}$ with a modulation depth of $\sim1\%$ under a 4~mW pump. The linear power dependence, mixing-ratio dependence, and detuning response consistently indicate \sakuma{a MIR and telecom photon-photon conversion} enabled by strong light--matter coupling. This approach establishes a polaritonic optomechanical platform for bridging disparate spectral regimes and provides a dissipation-engineered route toward hybrid MIR photonics and quantum transduction.
\end{abstract}

\section*{Keywords}
Vibrational polaritons, cavity optomechanics, MIR photonics

\section{Introduction}
Mid-infrared (MIR) photons, with wavelengths in the range of 3--20~$\mu\mathrm{m}$, offer a wide range of applications such as chemical sensing, molecular spectroscopy, and environmental monitoring. Advanced MIR photonic components, including narrow-linewidth quantum cascade lasers \cite{capasso2010high, yao2012mid}, sensitive photodetectors \cite{madejczyk2023research, long2017room, kajihara2013terahertz}, and high-quality photonic structures \cite{kazakov2024active, weng2017continuous}, have been developed owing to recent progress in semiconductor technologies. For many emerging applications, dynamic and high-speed control of MIR photons is highly desirable. 

\sakuma{Among the approaches proposed for the dynamic control of MIR photons, two particularly promising approaches are nonlinear optical frequency conversion and cavity optomechanics. In nonlinear optical frequency conversion, ultrafast polarization responses in nonlinear crystals enable direct frequency conversion between different bands. Telecom photons have been extensively coupled to visible and near-infrared photons using materials such as lithium niobate or III--V semiconductors via second- or third-order nonlinear optical effects \cite{li2022efficient, li2026pockels, kultavewuti2015low}. Although these approaches are highly effective in the ultraviolet to near-infrared regimes, extending them to the MIR is challenging due to material absorption and limited phase-matching conditions. 
Another promising approach is cavity optomechanics, where telecom photons and microwave-frequency signals are coupled via a mechanical mode \cite{aspelmeyer2014cavity}. By integrating an optomechanical cavity with a microwave circuit, telecom photons can be converted to microwave photons at the quantum level, providing a coherent interface between optical and microwave domains \cite{arnold2020converting,mirhosseini2020superconducting,holzgrafe2020cavity,brubaker2022optomechanical}. Although recent advances have enabled cavity optomechanics with phonon frequencies exceeding 10 GHz \cite{kharel2019high,xie2024sub,diamandi2025optomechanical}, direct extension toward the MIR regime remains challenging. These limitations motivate the exploration of alternative mechanisms that can mediate coherent interactions across widely separated spectral domains without relying on conventional direct nonlinear conversion or high-frequency nanomechanical modes.}

Here, we demonstrate a polaritonic optomechanical approach for controlling MIR photons using telecom photons in a polymer-based optical cavity. Rather than relying on direct photon–photon coupling, our platform exploits vibrational polaritons---hybrid states arising from strong coupling between MIR photons and  molecular vibrations. The ultrahigh-frequency molecular vibration can coherently interact with telecom photons through Raman-type optomechanical processes \cite{roelli2016molecular,chen2021continuous, jakob2023giant, jakob2025optomechanical, zhang2021addressing,zhang2013chemical}. This mechanism enables a molecular-vibration-mediated \sakuma{frequency conversion and dissipation engineering} across widely separated spectral regions. Importantly, polymers supporting strong light–-matter coupling in the MIR are typically transparent in the telecom band, offering a scalable platform for hybrid MIR photonics and MIR–telecom interfacing.

\section{Results and discussion}
Figure~1(a) shows a conceptual illustration of our setup based on an optical cavity that simultaneously supports telecom and MIR modes. Owing to dipole interaction, MIR cavity photons \sakuma{at $\omega_\mathrm{MIR}$} strongly couple with molecular vibrations \sakuma{at $\Omega_\mathrm{vib}$} to form vibrational polaritons when their frequencies are resonant \cite{long2015coherent, george2015liquid, xiang2018two, f2018theory}. In this regime, the MIR spectrum exhibits two resonance peaks corresponding to the upper and lower polariton branches, with a splitting given by twice the coupling rate, $2g$. 

To optomechanically control vibrational polaritons using telecom photons \sakuma{as a pump field}, two requirements must be satisfied for the cavity modes. The first requirement is that the dissipation rate of the telecom cavity, $\kappa_\mathrm{tel}$, is larger than the dissipation rate of molecular vibrations, $\gamma_\mathrm{vib}$. This condition ensures that the telecom photons can be \sakuma{coherently} eliminated, allowing the dynamics to be governed by the vibrational degrees of freedom. Because the dissipation rate of molecular vibrations is typically in a range of $0.1$ to $1$~THz for vibrational frequencies $\Omega_\mathrm{vib}$ of several tens of THz, achieving this condition requires a relatively low optical quality factor in the telecom band on an order of $10$--$100$.

The second requirement is a dual-cavity configuration in the telecom band due to the deep resolved-sideband regime, $\Omega_\mathrm{vib} \gg \kappa_\mathrm{tel}$. Since $\Omega_\mathrm{vib}\gg\kappa_\mathrm{tel}$, it becomes difficult to efficiently inject pump photons into \sakuma{the same cavity mode with the converted photon}. To overcome this limitation, we adopt a dual-cavity configuration in which two cavity modes are separated by a frequency close to $\Omega_\mathrm{vib}$ [see Fig.~1(b)]. 

By injecting a pump laser into the telecom-cavity mode, this configuration enables an efficient beam-splitter interaction between \sakuma{telecom photons in the Anti-Stokes (AS) cavity mode} and molecular vibrations, mediated by the cavity confinement \cite{aspelmeyer2014cavity}. \sakuma{This interaction corresponds to an AS scattering process, in which a vibrational excitation is coherently converted into a higher-frequency telecom photon. Such a beam-splitter interaction is well known in cavity optomechanics to cool the mechanical motion, leading to a reduction in the vibrational amplitude and an effective shortening of its lifetime, i.e. the effective dissipation enhancement}.

\begin{figure}
  \centering
  \includegraphics[width=12cm]{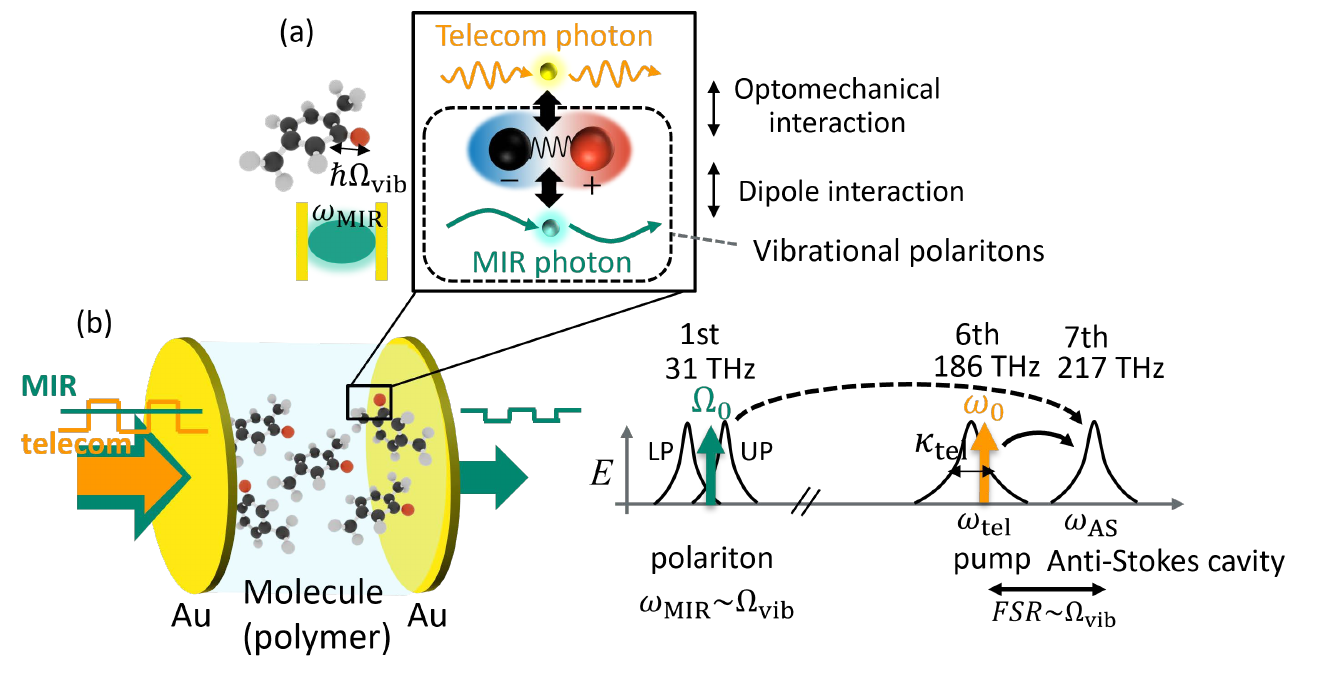}
  \caption{
    (a) Schematic of a Fabry–Pérot optical cavity. An MIR pulsed laser and an intensity-modulated pump laser are injected along the same optical path and induce dipole and optomechanical interactions. The MIR photons resonantly couple with molecular vibrations to form vibrational polaritons, while the telecom pump photons cool the molecular vibrations through a beam-splitter-like interaction. 
    (b) Schematic of resonant cavity modes. Vibrational polaritons are hybrid states formed by the strong coupling between the cavity photon mode at $\omega_\mathrm{MIR}$ and the molecular vibrational mode $\Omega_\mathrm{vib}$. The molecular vibrations are controlled by a parametric process induced between pump photons ($\omega_0$) and \sakuma{Anti-Stokes (AS) cavity mode $\omega_\mathrm{AS}$}}.
  \label{fgr:one}
\end{figure} 

To demonstrate our concept, \sakuma{we employ a Fabry--Pérot optical cavity consisting of Au mirrors and a poly(chloro-\textit{p}-xylylene) (Parylene-C) spacer layer}, fabricated by thermal evaporation of Au and chemical vapor deposition of Parylene-C. Parylene-C is transparent in the telecom band \cite{guggenheim2021visible} and exhibits a C--Cl stretching vibrational mode around $10~\mu\mathrm{m}$ \cite{herman2020determination}, making it a suitable polymer for implementing our scheme. From a reference single layer of Parylene-C deposited under the same conditions, we estimate $\Omega_\mathrm{vib}/2\pi = 31.5$~THz (corresponding to a wavelength of $9.5~\mu\mathrm{m}$) and $\gamma_\mathrm{vib}/2\pi =0.5$~THz by scanning the MIR frequency [see Supplemental Information].

To satisfy the first requirement, the thickness of the Au mirrors is designed to be $20$~nm, resulting in a telecom cavity dissipation rate $\kappa_\mathrm{tel}/2\pi =1.0$~THz, which exceeds $\gamma_\mathrm{vib}$ as confirmed by scanning the telecom laser frequency. This ensures the coherent elimination of the telecom photons. To fulfill the second requirement, the thickness of the Parylene-C layer is set to $3.2~\mu\mathrm{m}$ such that the cavity free spectral range in the telecom band is nearly resonant with $\Omega_\mathrm{vib}$. Figure~2(a) shows reflection spectra of a telecom laser from 1.36 to 1.63~$\mathrm{\mu m}$ (184 to 220~THz) at an incident angle of $40^\circ$. These two dips correspond to the sixth and seventh resonant modes of the cavity. Owing to material dispersion, the sixth optical mode (frequency $\sim 186$~THz, wavelength $\sim 1.6~\mu\mathrm{m}$) predominantly couples to the adjacent lower- and higher-order modes. The coupling to \sakuma{the adjacent higher-order seventh modes} is stronger than that to \sakuma{the lower-order fifth mode} due to reduced optical absorption at higher frequencies. As a result, the system naturally operates in a red-detuned pumping configuration, favoring a beam-splitter-type interaction and \sakuma{vibration suppression, thereby improving the efficiency of the dissipation enhancement in our scheme.} A model including the effects of both higher- and lower-order modes is discussed in the Supplemental Information.

First, we excited vibrational polaritons using a quantum cascade laser (QCL) covering a wavelength range of 5.5--11.0~$\mathrm{\mu m}$. A pulsed laser with a repetition rate of 38~kHz was injected, and the transmitted power through the Fabry--Pérot optical cavity was detected using a mercury cadmium telluride (MCT) detector. Background thermal signals were removed by lock-in detection at the MIR pulse frequency. The detailed measurement setup is illustrated in Fig.~S3. Figure~2(b) shows a colormap of the cavity transmission spectra, where the color scale represents the transmitted power, plotted as a function of the incident angle (x-axis) and wavelength (y-axis). \sakuma{Changing the incident angle modifies the cavity effective length, enabling the tuning of cavity frequency.} The coupling between the C--Cl stretching vibrational mode and the cavity resonance results in an avoided crossing, forming the lower polariton (LP) and upper polariton (UP) branches. At an incident angle of $25^\circ$, the UP appears at a wavelength of 9.5~$\mathrm{\mu m}$ (frequency: 31.5~THz), exhibiting a vibration-dominant character, while the LP appears at 10.2~$\mathrm{\mu m}$ (29.4~THz), exhibiting a photon-dominant character. 

At an incident angle of $40^\circ$ (gray dashed line in Fig.~2(c)), the cavity becomes resonant with the C--Cl stretching vibration, resulting in an equal mixing of photonic and vibrational components in the UP and LP branches. The splitting between the UP and LP is 1.0~THz, corresponding to a coupling strength of $g=0.5$~THz, which exceeds the dissipation rates of both modes, confirming the strong-coupling regime. 
To demonstrate optomechanical control of MIR photons via telecom photons, we injected a pump laser into \sakuma{the sixth mode of} the Fabry--Pérot cavity along the same optical axis as the MIR beam. The pump intensity was modulated using an optical chopper operating at 21~Hz, and the resulting change in the transmitted MIR intensity was measured with a lock-in amplifier. Figure~2(c) shows the change in transmitted MIR power as a function of wavelength and MIR detuning under a pump power of 4~mW. It is evident that the MIR photons are attenuated depending on the MIR frequency, \sakuma{with pronounced attenuation occurring at the wavelengths corresponding to the upper and lower polariton resonances}. This behavior can be understood as follows: the vibrational component of the polariton \sakuma{is parametrically converted to telecom photons in the AS cavity} via the optomechanical interaction, which in turn reduces the equilibrium population of the MIR photons. 

\begin{figure}
  \centering
  \includegraphics[width=12cm]{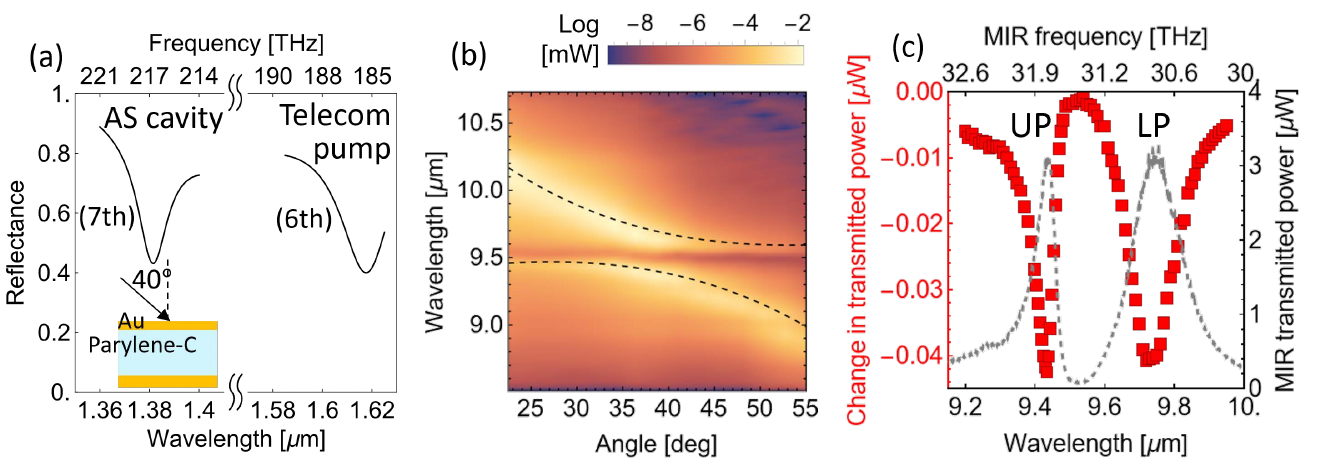}
  \caption{
    (a) Optical reflection spectrum in the telecom band at an incident angle of $40^\circ$. Two different lasers operating around 1600 nm and 1380 nm were used to probe the resonance modes corresponding to the telecom (sixth-order) and AS (seventh-order) cavity modes, respectively. The spectra shown in this figure represent fitted resonance lineshapes obtained from the measured reflection data. The figure inset shows a schematic of the cavity containing parylene-C.
    (b) Colormap of the transmission spectra of the Fabry–Pérot cavity containing Parylene-C as a function of the incident angle (x-axis) and wavelength (y-axis). The transmission spectra were measured at intervals of $2.5^\circ$. 
    (c) Change in the MIR transmitted power using a pump input at a frequency of 186 THz (red markers). The gray dashed line shows the MIR transmission spectrum at an incident angle of $40^\circ$.
    }
  \label{fgr:two}
\end{figure}

\sakuma{Being analogous to the sideband cooling in optomechanical systems, the suppressed transmission can be regarded as the polariton cooling induced by the telecom pump.} This operation enables control of the MIR photons via two approaches: (1) pump power and (2) modulating the incident angle to vary the mixing ratio between MIR photons and molecular vibrations. These control parameters provide enhanced tunability and flexibility, enabling precise and multidimensional control of MIR photonic states via telecom photons.

To investigate the first control parameter, we increased pump power up to 3~mW and measured the change in MIR attenuated power. The MIR attenuated power increases linearly with the pump power at both the LP and UP peaks, as shown in Fig.~3(a). From a linear fit to the pump-power dependence, we estimate the attenuation rate to be approximately $0.3\%$ per $1$~mW of pump input. This linear dependence reflects that the change in the MIR output power arises from optomechanical three-wave mixing \sakuma{among vibrational polaritons, telecom photons, and AS-cavity photons}, as expected from our model. The observed attenuation rate allows us to estimate the Raman activity of parylene-C using a model that incorporates both dipole and dispersive interactions between photons and vibrations. The estimated Raman activity is on the order of \SI{10}{\angstrom^4\per amu}, which is within the typical range obtained in density functional theory (DFT)-based Raman analyses of organic molecules and polymeric vibrational modes \cite{kinayturk2023spectroscopic, virdi2003ab}, thereby supporting the validity and consistency of the present results [see Supplemental Information].

To explore the second control parameter---the mixing ratio between MIR photons and vibrations---we varied the incident angle of the MIR laser. Figure~3(b) shows the evolution of the transmission power ratio between the lower polariton (LP) and upper polariton (UP) peaks as a function of incident angle and cavity detuning ($\Delta_\mathrm{cav}=\omega_\mathrm{MIR}-\Omega_\mathrm{vib}$). As the incident angle increases, the LP-to-UP ratio decreases. This trend indicates that the UP branch becomes increasingly photon-like, while the LP branch acquires a more vibration-like character. Figure~3(c) summarizes the corresponding changes in the maximum attenuation at the LP and UP peaks with a 1~mW pump input, plotted as a function of incident angle and cavity detuning. At zero cavity detuning, the attenuation at the LP and UP peaks is comparable, reflecting the equal mixing of photonic and vibrational components in both branches. \sakuma{For positive cavity detuning, the UP peak exhibits a larger attenuation than the LP peak. This asymmetric cavity-detuning dependence plays a crucial role in elucidating the parametric processes that cause polariton cooling.}

There are two possible scenarios that could cause polariton cooling. One is that the \sakuma{vibrational} component is parametrically cooled by the pump photons, leading to the polariton cooling. The other is that the cooling of the photonic component causes polariton cooling. The angle dependence in Figure 3(c) can be well explained by assuming the former case: the cooling through the \sakuma{vibrational} component. As the incident angle increases, the attenuation decreases for both polariton branches. \sakuma{Importantly, the decrease is more pronounced for the LP branch than for the UP branch. This asymmetry arises from vibration cooling}: for the UP branch, the increasing photonic character reduces the influence of vibration cooling, whereas for the LP branch, the increasing vibrational character suppresses the transduction of vibration cooling into the MIR output. Consequently, the modulation of the transmitted MIR power decreases for both branches with increasing incident angle. In contrast, a photon-cooling scenario predicts stronger cooling for the photon-like branch (see Supplemental Information), which is inconsistent with the experimental observations. These results therefore support that the observed effect is governed by vibration cooling.


\sakuma{The branch-dependent cooling effect} indicates that MIR photons corresponding to either the LP or UP branch can be selectively attenuated by controlling the cavity detuning. Tuning the incident angle enables wavelength-selective control of MIR photons associated with the polariton branches. The attenuation behavior strongly depends on the cavity dissipation rates. With a lower cavity dissipation rate, stronger optical gain is preferentially applied to the branch with a higher photonic character. As a result, the attenuation rates of the vibration-like and photon-like branches become significantly different, enabling more pronounced wavelength-selective control of MIR photons (see Supplemental Information). However, as discussed above, excessively reducing the cavity dissipation rates suppresses the optical gain transferred to molecular vibrations. 

\begin{figure}
  \centering
  \includegraphics[width=12cm]{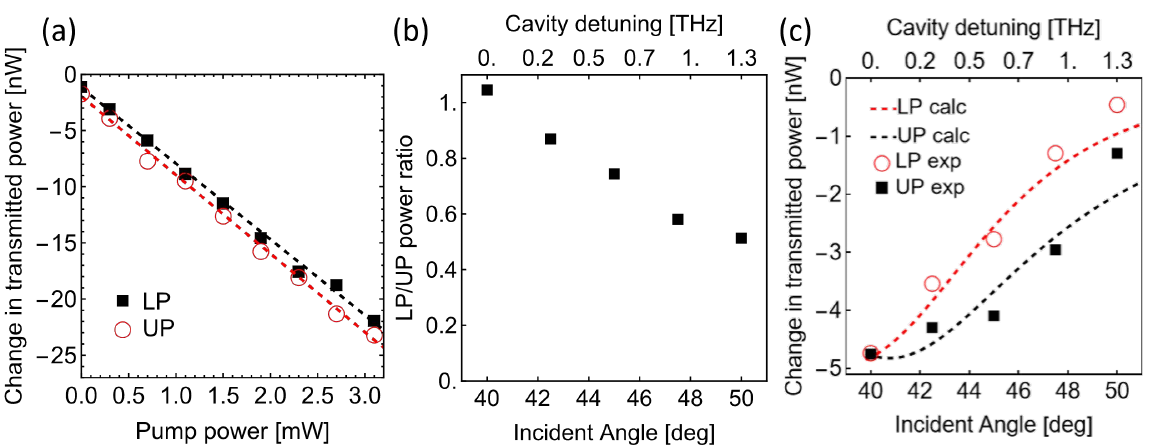}
  \caption{
    (a) Change in the transmitted MIR power with increasing pump power at the LP (square markers) and UP (circle markers) peaks. Dashed lines indicate best-fit lines for LP and UP markers.
    (b) Change in LP and UP power ratio at different incident angles.
    (c) Attenuated MIR power as a function of incident angle. Cavity detuning at upper x-axis is calculated by $\Delta=\omega_\mathrm{MIR}-\Omega_\mathrm{vib}$. Red and black markers: experimentally obtained MIR power at the LP and UP peaks. Red and black dashed lines: calculated MIR power at the LP and UP peaks. The pump detuning was set to zero.
     }
  \label{fgr:three}
\end{figure}

To maximize the attenuation efficiency, the pump detuning must be optimized for each MIR wavelength. Figure~4(a) shows the change in the attenuated MIR power at a wavelength of $9.77~\mu\mathrm{m}$ (frequency: 30.7 THz) and an incident angle of $44^\circ$ as a function of the pump detuning. The attenuation is maximized at a finite pump detuning, indicating that the optimal condition is shifted from the cavity resonance due to the parametric interaction mediated by molecular vibrations.

To further investigate this behavior, we experimentally measured the pump-detuning dependence at each MIR frequency around the LP ($\sim$31.0~THz) and UP ($\sim$31.9~THz) peaks. The color map in Fig.~4(b) shows the change in the attenuated MIR power as a function of MIR frequency and pump detuning. The white markers indicate the pump detuning corresponding to the maximum attenuation power at each MIR frequency. Interestingly, the optimal detuning exhibits a sawtooth-like dependence in both the LP and UP branches. This characteristic behavior can be qualitatively understood from energy conservation among the telecom pump photons, scattered photons, and molecular vibrations involved in the parametric process. As the MIR frequency changes, the frequency-matching condition required for efficient parametric \sakuma{cooling} is correspondingly modified, resulting in the observed sawtooth-like dependence. The clear emergence of this behavior indicates that the \sakuma{cooling} originates from the vibration-mediated parametric interaction rather than from direct optical coupling. Calculated attenuation characteristics based on energy conservation between vibrational and telecom-photon interactions reproduced a qualitatively similar sawtooth-like trend (Fig.~S7). A more comprehensive understanding could be obtained by extending the model to include collective vibrational-polariton effects with finite linewidths, inhomogeneous broadening, and possible dark-state contributions.

\begin{figure}
  \centering
  \includegraphics[width=8cm]{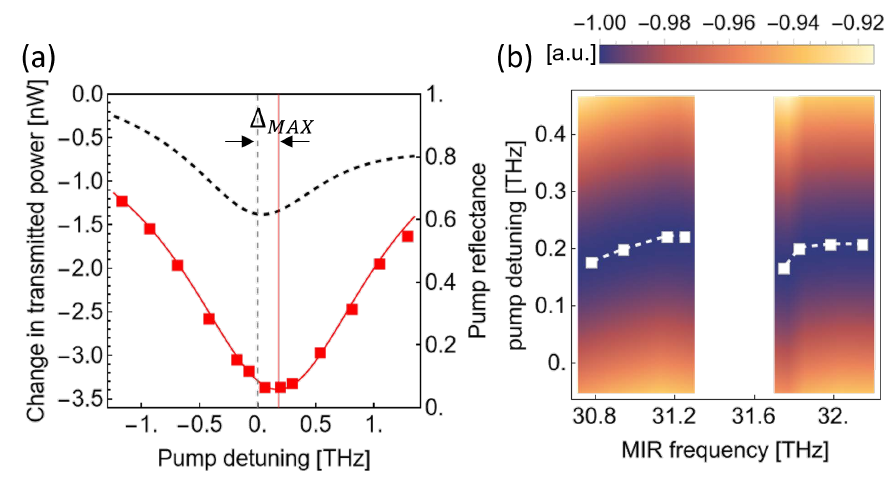}
  \caption{
    (a) Change in the transmitted MIR power (square markers) and reflection spectrum of the telecom pump input (dashed line) as a function of pump detuning at an incident angle of $44^\circ$. The solid line shows the fitting curve. 
    (b) Colormap of change in the transmitted power as a function of MIR frequency (x-axis) and pump detuning (y-axis). The square markers are pump detunings corresponding to the maximum MIR attenuated power. \sakuma{The frequency range between 31.3 and 31.7 THz is omitted in Fig.~4(b), where the MIR signal is strongly suppressed in the dark-mode region between the LP and UP resonances.}
   }
  \label{fgr:four}
\end{figure}


To further improve the MIR attenuation rate of $0.3\%/\mathrm{mW}$ observed in the present system, both the photon-vibration coupling strength and the optical confinement should be enhanced. Using materials with a higher Raman activity or a larger vibrational dipole moment directly strengthens the optomechanical and dipole interactions, while reducing the optical mode volume increases the intracavity field intensity. 
For example, increasing the Raman activity by one order of magnitude ($\sim300$ \si{\angstrom^4\per{amu}}) together with reducing the optical mode volume to $\sim0.1~\mathrm{\mu m^3}$ (10\% of the current value) is expected to yield an attenuation rate of approximately 50\%/mW. However, large Raman activity and strong vibrational dipole moments are generally difficult to achieve simultaneously in a single material system, making careful material selection essential. 
Such parameter regimes are, in principle, accessible using nanofabricated resonators with tightly confined electromagnetic fields \cite{arul2022giant, hu2025plasmonic} and molecular systems with large Raman activity or strong vibrational dipoles, such as Rhodamine 6G (R6G) \cite{watanabe2005dft}, disiloxane \cite{carteret2007ab}, and poly(methyl methacrylate) (PMMA) \cite{virdi2003ab}. These considerations suggest a realistic pathway toward substantially enhanced MIR attenuation performance.

At the same time, the cavity dissipation rate should not be reduced excessively: although a higher cavity $Q$ factor strengthens the optical field buildup, it also decreases $\kappa_\mathrm{tel}$ and can drive the system away from the regime favorable for vibration cooling. Therefore, the optimal device should be designed not for the highest possible cavity $Q$, but for an intermediate regime in which strong field confinement is maintained while the inequality $\kappa_\mathrm{tel} \gtrsim \gamma_\mathrm{vib}$ is still satisfied. 

Moreover, beyond coherent classical control of MIR photons, our system provides a promising platform for quantum light--matter interfaces. Since the demonstrated operation enables controlled and coherent extraction of vibration excitations hybridized with MIR photons in the cavity, it can be interpreted as a form of measurement of the vibrational component of the vibrational polariton. Furthermore, the interaction mediated by telecom photons suggests the possibility of accessing the system from the telecom band, potentially enabling heralded detection schemes. Such capabilities could open new avenues for quantum operations, including weak-measurement-based state control \cite{shomroni2019optical}, non-Gaussian state generation \cite{pepper2012optomechanical}, and other forms of quantum state engineering. In addition, the vibrational frequency in our system lies in the tens of terahertz range, where the thermal phonon occupation is intrinsically suppressed even at room temperature, enabling access to phononic states close to the quantum ground state without requiring cryogenic environments. Furthermore, the optomechanical approach to controlling vibrational polaritons presented here may extend the rich toolbox of nonlinear control and many-body dynamics developed for conventional solid-state mechanical resonators \cite{heinrich2011collective,slim2024optomechanical,asano2025synthesized} to phononic excitations in matter. Such capabilities could establish a versatile platform at the intersection of quantum physics, condensed-matter physics, and nonlinear dynamics, enabling new opportunities for coherent phonon engineering and collective vibrational phenomena in the MIR regime.

\section{Conclusion}

In conclusion, we have demonstrated optomechanical control of MIR photons mediated by vibrational polaritons, achieving telecom-driven \sakuma{dissipation enhancement} of MIR photons at $9.5~\mu\mathrm{m}$ with a modulation depth of $\sim1\%$ under a $4$ mW pump. The linear pump-power scaling and characteristic detuning response consistently indicate \sakuma{an optomechanical parametric conversion} between telecom photons and molecular vibrations. This approach establishes a polaritonic optomechanical platform for all-optical control of MIR photons, providing a pathway to interface MIR photonics with mature telecom-band technologies. Furthermore, leveraging vibrational degrees of freedom offers opportunities for hybrid photonic functionalities and potential quantum transduction in the MIR regime.

\section{Data Availability}
All of the data supporting the findings of this study are presented in the Results section and the Supporting Information and are available from the corresponding author upon reasonable request.

\section{Terms \& Conditions}
The authors declare no competing financial interest.

\section{Acknowledgements}
The authors acknowledge Hisashi Sumikura for his valuable assistance with the experimental setup. This work was supported in part by JSPS KAKENHI (23H05463, 26K01424).

\printbibliography
 
\end{document}